\documentclass[american,aps,prl,reprint,groupedaddress,twocolumn,showpacs]{revtex4-1} 
\usepackage[unicode=true,pdfusetitle, bookmarks=true,bookmarksnumbered=false,bookmarksopen=false, breaklinks=false,pdfborder={0 0 0},backref=false,colorlinks=false] {hyperref}
\hypersetup{ colorlinks,linkcolor=myurlcolor,citecolor=myurlcolor,urlcolor=myurlcolor}
\usepackage{braket,colortbl,cleveref,amsthm,amsmath,amssymb,txfonts}
\definecolor{myurlcolor}{rgb}{0,0,0.7}
\usepackage{graphics,graphicx}
\usepackage{color}
\usepackage{enumerate}
\usepackage{colortbl}
\usepackage{subfigure}
\usepackage{graphicx}
\usepackage{ragged2e}
\usepackage{amsmath}
\usepackage{graphicx}
\usepackage[utf8x]{inputenc}
\usepackage{color}
\usepackage{amsmath}
\usepackage{tikz}
\usetikzlibrary{shapes.geometric, arrows}
\usepackage{latexsym}
\usepackage{amssymb}
\usepackage{amsthm}
\usepackage{bm}
\usepackage{graphics,epstopdf}
\usepackage{color}\usepackage{amsmath}

\usepackage[usenames,dvipsnames,svgnames]{pstricks}
\usepackage{epsfig}
\usepackage{pst-grad} 
\usepackage{pst-plot} 
\tikzstyle{startstop} = [rectangle, rounded corners, minimum width=3cm, minimum height=1cm,text centered, draw=black, fill=red!30]

\tikzstyle{env}=[circle,  ball color = green!20, minimum size= 80mm]
\tikzstyle{central}=[circle, ball color = red!100, minimum size=8mm]
\tikzstyle{bath}=[circle, ball color =blue!75, minimum size=4mm]

\theoremstyle{plain}

\def\bea{\begin{eqnarray}}
\def\eea{\end{eqnarray}}
\def\ba{\begin{array}}
\def\ea{\end{array}}

\def\beq{\begin{equation}}
\def\eeq{\end{equation}}

\begin{document}

\title{ Superposition of causal order enables perfect quantum teleportation with very noisy singlets}

\author{Chiranjib Mukhopadhyay}
\email{chiranjibmukhopadhyay@hri.res.in}
\author{Arun Kumar Pati}
\email{akpati@hri.res.in}
\affiliation{Quantum Information and Computation Group, Harish-Chandra Research Institute, Homi Bhabha National Institute, Allahabad 211019, India }


\begin{abstract}
\noindent  
We show that the fidelity of the standard quantum teleportation protocol, which utilizes an impure resource state, applied successively, can be significantly improved, when used in conjunction with a quantum switch. In particular, we find that for two such teleportation channels conjugated with the superposition of causal order, teleportation fidelity beyond the classical threshold is achieved for significantly larger noise than would be possible conventionally. One can even make the effective teleportation channel perfectly faithful for very large noise. We also discuss the generalization of our scheme for more than two pathways, and define a figure of merit in that context. 


\end{abstract}

\maketitle

\emph{Introduction-} Teleportation of unknown quantum states \citep{teleport, bouwmeester, teleport_prob} is a cornerstone of quantum information science. However,  perfect implementation of the standard teleportation protocol \citep{teleport} requires singlets, which are highly fragile. Hence, in practical situations, imperfect singlets \citep{horodecki, bouwen_bose} must be considered, where the degree to which the resource state deviates from a perfect singlet, governs the degradation in the fidelity of teleportation. Eventually, if the imperfection grows beyond a certain threshold, the resulting fidelity can be met or exceeded through classical means, which indicates that the standard teleportation protocol no longer furnishes any quantum advantage. In this letter, we show that it is possible to probabilistically retain such a quantum advantage even if the resource state significantly differs from a perfect singlet, if the sender and the receiver have access to a quantum switch \citep{branciard,oreshkov,brukner,ebler,chiribella_PRA,manik, Rubino, Expt, guerin}. In fact, we show that a higher amount of imperfection may actually turn out to be more helpful towards quantum teleportation. Quantum switches contain a control qubit, whose basis elements map to a distinct sequential orderings of quantum channnels each. If the control qubit is initialized in a superposition of these basis elements, then the physical scenario is an example of processes with a superposition of causal order \citep{oreshkov,brukner, chiribella_review}. Such processes have been recently utilized to offer a reduction in query complexity \citep{araujo}, enhancement of classical capacity of quantum channels \citep{ebler, branciard, manik}, and improvement in steady state quantum thermometry \citep{our_thermometry} among other tasks. The present work fits into this paradigm as another explicit example where the superposition of causal order spawns a definite operational advantage.


\emph{Teleportation Protocol as a generalized depolarizing channel -} The goal of the standard teleportation protocol is to transfer the information of an unknown qubit to a  different location without physically sending a qubit. This is accomplished through using a singlet shared between parties at the two locations, that is, an ebit, and two classical bits of information. If the shared state is indeed a singlet $|\psi^{-}\rangle = \frac{1}{\sqrt{2}} \left( |01\rangle - |10\rangle \right)$, the state is teleported with perfect fidelity.  However, if the shared bipartite state, say $\chi$, is an arbitrary one, then implementing the standard teleportation channel protocol can be shown to be equivalent to a generalized depolarizing channel \citep{bouwen_bose}  $\Lambda$ on the state $\rho$ that we wish to send, that is, $\Lambda [\rho] = \sum_{i = 0}^{3} p_{i} \sigma_{i} \rho \sigma_{i}$. The weights $\lbrace p_i \rbrace$ of the generalized depolarizing channel are the overlaps of the shared bipartite state $\chi$ with the elements of the Bell basis. In particular, $p_0$ is the singlet fraction of the shared resource state $\chi$. Throughout the letter, we shall assume the simplification $p_1 = p_2 = p_3 = p$, and $p_0 = 1 - 3p$. This places the constraint $0 \leq p \leq 1/3$ on the value of $p$.

\begin{figure}
\includegraphics[scale=0.25]{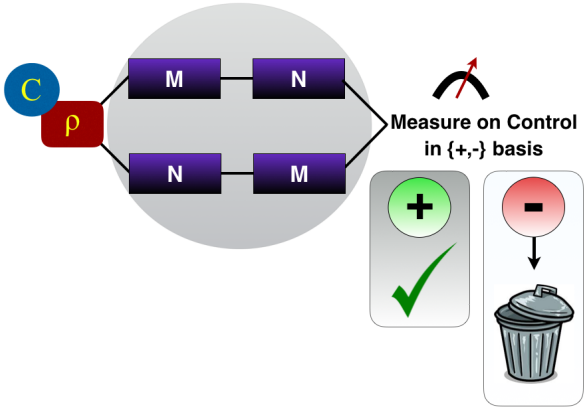}
\caption{Schematic of the protocol. Identical teleportation channels M and N are applied In one path as first M then N, in the other path as first N then M.  These paths are then superposed via the control which acts as the quantum switch. }
\label{protocol_fig}
\end{figure}

\emph{Scheme}- We assume two identical teleportation channels are applied back to back, and the ordering between them is controlled by a quantum switch. If the control qubit of the quantum switch is at a state $|0\rangle$, then one ordering is unambiguously followed, if the control qubit is at a state $|1\rangle$, then the reverse ordering is unambiguously followed. Finally, the control qubit of the quantum switch is measured in the Hadamard basis. If $\lbrace K_i \rangle$ is the set of Kraus operators of one of the two identical channels, then the correlated output of the final state may be shown to be given by $\sum_{i}\sum_{j} W_{ij} (\rho \otimes \rho_{c})W_{ij}^{\dagger}$, where $W_{ij} = K_{i} K_{j} \otimes |0\rangle \langle 0| + K_{j} K_{i} \otimes |1\rangle \langle 1|$. The final joint state hides information about the input qubit in the correlations between the two paths, which is revealed in the final state of the system, when the measurement in the Hadamard basis is performed on the control qubit. If the outcome $|+\rangle = \frac{1}{\sqrt{2}} \left(|0\rangle + |1\rangle \right)$ is obtained as a result of measurement, then the protocol succeeds. If the outcome $|-\rangle$ is obtained, the protocol fails. A schematic of the protocol is given in Fig. \ref{protocol_fig}. If the control qubit is initialized in the state $\sqrt{q} |0\rangle + \sqrt{1-q} |1\rangle$, then the unnormalized post-measurement state of the system, conditioned by outcome $\pm$ of the measurement on the control qubit, is given by 
\begin{widetext}
\begin{equation}
\rho_{out}^{\pm} = \sum_{i,j=0}^{3} \frac{p_i p_j}{2} \left[ q \sigma_{i} \sigma_{j} \rho \sigma_{j} \sigma_{i} \pm \sqrt{q(1-q)} \sigma_i \sigma_j \rho \sigma_i \sigma_j \pm  \sqrt{q(1-q)} \sigma_j \sigma_i \rho \sigma_j \sigma_i + (1-q) \sigma_j \sigma_i \rho \sigma_i \sigma_j \right]
\label{mother_formula}
\end{equation}
\end{widetext}

The teleportation channel considered here, when applied once, leads to a teleportation fidelity of $\mathcal{F}_1 = 1- 2p$, and when applied twice in succession, i.e., without the quantum switch, leads to a teleportation fidelity of $\mathcal{F}_2 = 1- 4p+ 8p^2.$ However, the teleportation fidelity has to exceed $2/3$ to demonstrate any quantum advantage over maximal state estimation fidelity. By this condition, this channel only allows  for a quantum advantage for $p \in [0, 1/6]$ , when applied once. When applied sequentially twice, the range of noise $p$ for which quantum advantage is gained shrinks further to $p \in [0, \frac{1}{4} (1- \frac{1}{\sqrt{3}})]$. That is, if the noise $p$ exceeds $\approx 0.1057$, then it is possible to come up with a better fidelity using solely classical means, than the standard teleportation protocol. The main result in this work is to show that, it is possible using a quantum switch to achieve a quantum advantage beyond this level of noise. 

\noindent \textbf{Result  I } : \emph{ For the protocol desribed above, when the measurement on the control state yields the outcome $+$, the standard teleportation protocol confers a genuine quantum advantage if one of the two following conditions is met}
\begin{enumerate}[(i)]
\item $0 \leq p < \frac{3 (1+ 2 \mu) - \sqrt{3} \sqrt{1+ 2 \mu}}{12 (1+ 3 \mu)}$,
\item $ \frac{3 (1+ 2 \mu) + \sqrt{3} \sqrt{1+ 2 \mu}}{12 (1+ 3 \mu)} < p \leq \frac{1}{3}$, \emph{where $\mu = \sqrt{q(1-q)}$.}
\end{enumerate}

\noindent \textbf{Proof -} From \eqref{mother_formula}, adequately normalizing the expression for post measurement state $\rho_{\text{out}}^{+}$, and considering an arbitrary input state $\rho$, the expression for fidelity  between the input and output states is given by $\mathcal{F} (\rho, \rho_{\text{out}}^{+}) = \text{Tr}  (\rho \rho_{\text{out}}^{+}) + 2 \sqrt{\text{det} (\rho) \text{det} (\rho_{\text{out}}^{+})}  =  \frac{1+ 2\mu - p (4 + 8 \mu) + 8 p^2 (1+ \mu)}{1 + 2\mu (1 - 12 p^2)}$. The first equality follows from the definition of fidelity in the qubit case \citep{fid_det}. Now, the fidelity must be greater than $2/3$ in order to confer genuine quantum advantage. Plugging this condition into the expression for fidelity leads to the following quadratic inequality in $p$, \[ \frac{1 + 2\mu}{3} - 4 (1+ 2 \mu) p + 8 (1+ 3 \mu) p^2 > 0.\] Solving this inequality yields the result above. \qed

\begin{figure}
\includegraphics[scale=0.25]{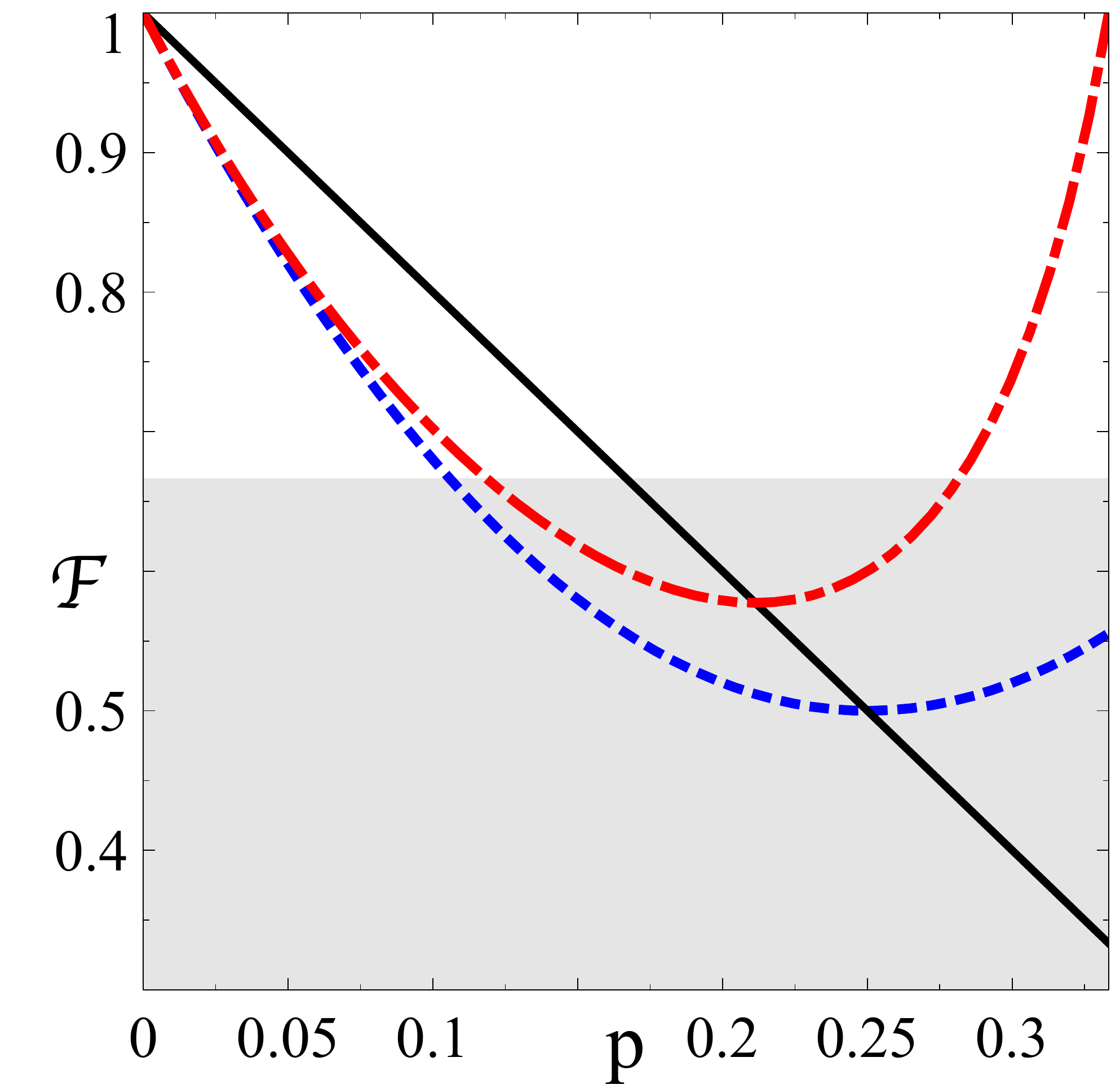}
\caption{Teleportation fidelity for a single teleportation channel (black solid line), two consecutive teleportation channels applied back to back (blue dotted line), and two teleportation channels, whose pathways are controlled by the control qubit at $\frac{1}{\sqrt{2}}\left( |0\rangle + |1\rangle \right)$, and the outcome of the measurement on the control qubit being $+$. The grey region is where there is no quantum advantage, i.e., teleportation fidelity is less than $2/3$.}
\label{2d_fig}
\end{figure}

\begin{figure}
\includegraphics[scale=0.2]{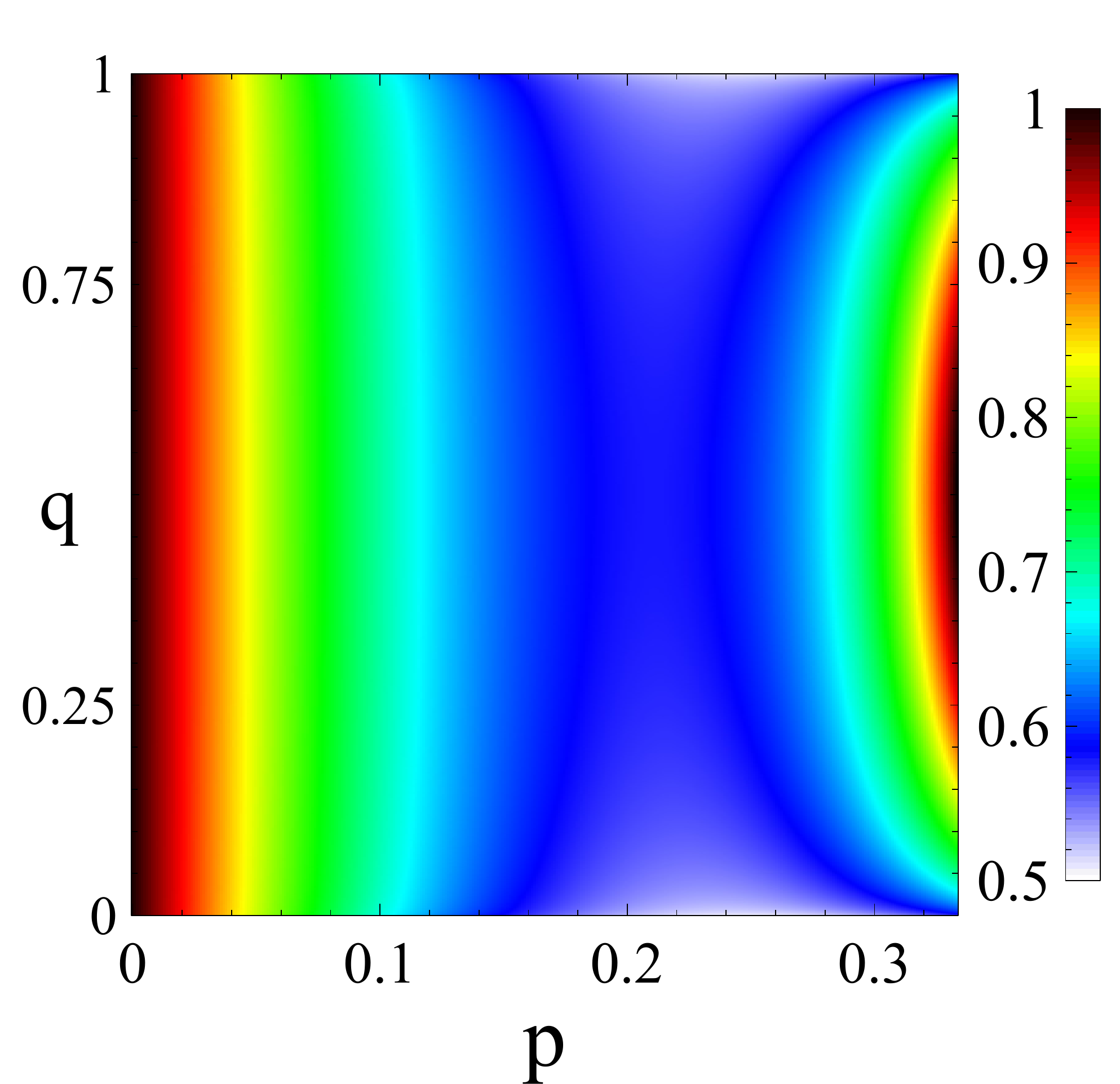}
\includegraphics[scale=0.19]{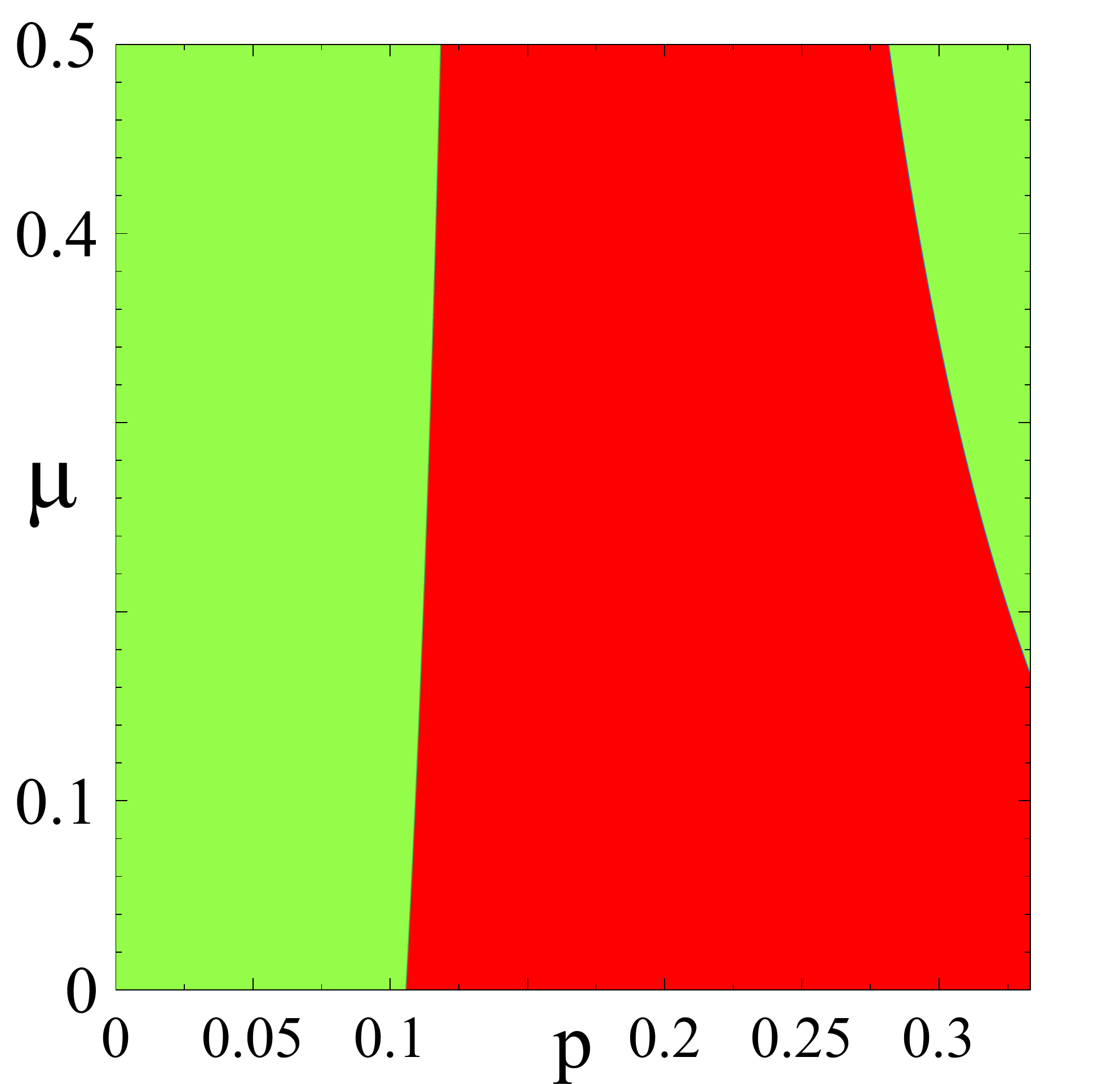}
\caption{\emph{Left:} dependence of teleportation fidelity on probabiliy $p$ for the isotropic noise model, and the superposition parameter $q$ of the quantum switch. \emph{Right:} The dependence of noise $p$ on the superposition parameter $\mu = \sqrt{q(1-q)}$ for which it is possible (green region) or impossible (red region) to generate teleportation fidelity greater than $2/3$.}
\label{3d_fig}
\end{figure}

\noindent As an illustration of the result above, Fig. \ref{2d_fig} depicts the situation for the control qubit initialized in the maximally coherent state $|+\rangle$. For any non-zero value of $\mu$, the upper bound to the first region of $p$ in the above equation is larger than $0.1057$. Even more intriguingly, the second region of $p$, when it exists, for which the teleportation fidelity is larger than $2/3$, does not have any previous analog. It may be mentioned here, that in order for this second region to be non-null, the superposition parameter of the quantum switch $\mu$ must satisfy the constraint 
\begin{equation}
\frac{3 (1+ 2 \mu) + \sqrt{3} \sqrt{1+ 2 \mu}}{12 (1+ 3 \mu)}  < \frac{1}{3}
\end{equation}
\noindent Solving the above inequality, one obtains the result that the superposition parameter $\mu$ must be greater than the threshold value of $1/6$, as depicted in Fig. \ref{3d_fig}, in order for the second region, which indicates a revival of quantum advantage in teleportation under very noisy channels, to exist. In the second region, the teleportation fidelity increases monotonically with the strength of the noise $p$. This justifies the assertion that large noise helps rather than hinders quantum teleportation. 

\emph{Lossless transmission with two channels -}  For any physical system obeying the superposition principle, destructive interference is always a possibility. Thus, it is natural to wonder about the possibility whether the two pathways for teleportation may somehow be made to destructively interfere, rendering the effective channel to be lossless. For the above noise model, when the measurement on the control qubit renders the result $+$, the teleportation is perfect when $\mu = 1$, i.e., the control qubit is in equal superposition of both the paths, and $p = 0 $ or $p= 1/3$. While the former is the trivial case of teleportation using a perfect singlet, the latter is especially noteworthy, as here we have the maximal noise in the resource state, yet the qubit is faithfully teleported. We must reiterate here that the protocol is inherently probabilistic in nature, contingent upon obtaining the $+$ outcome for the measurement on the control qubit. 

\begin{figure}
\includegraphics[scale=0.35]{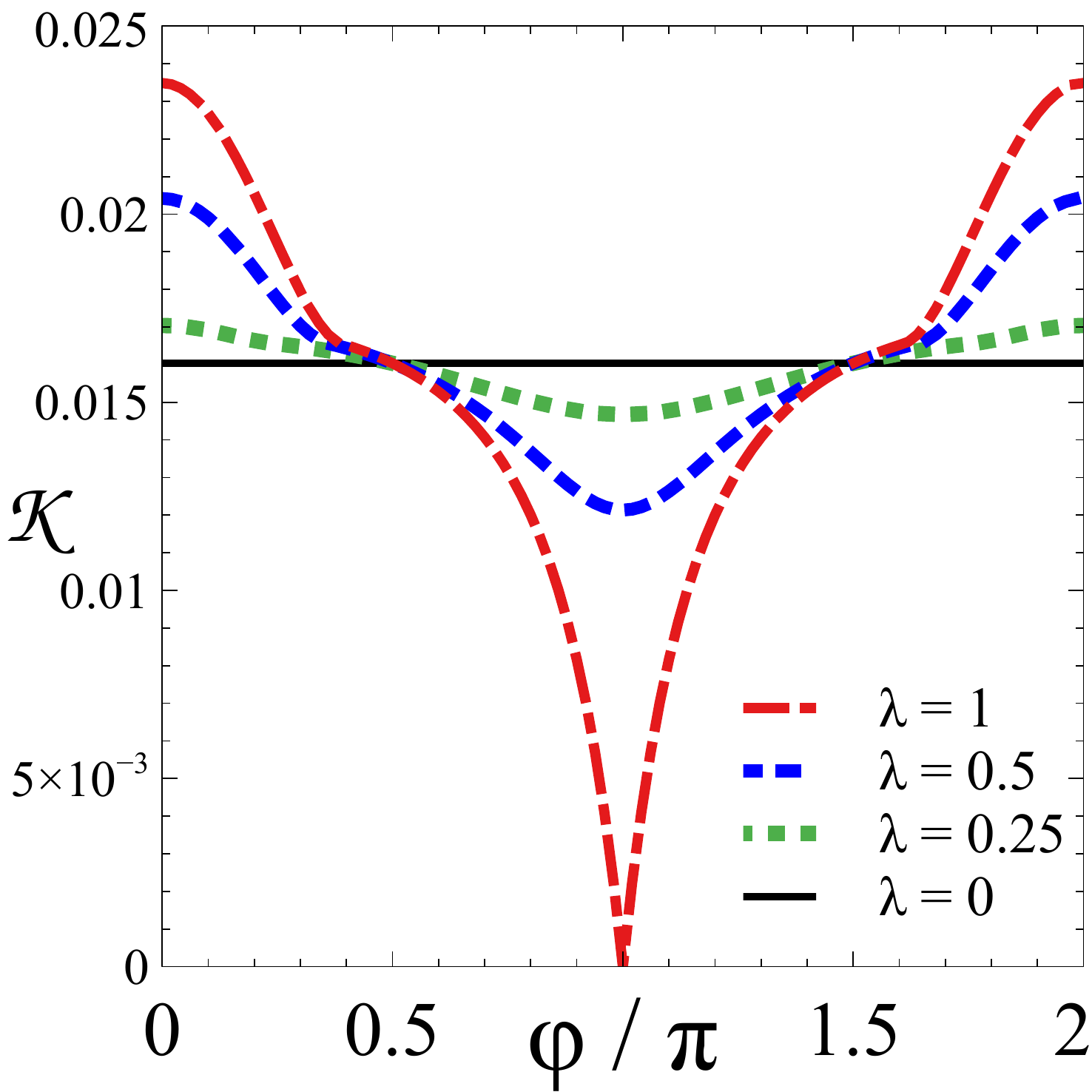}
\caption{Figure of merit $\mathcal{K}$ vs. measurement outcome $|0\rangle + \lambda e^{i \phi} |1\rangle$ on the control part of the output, over which the outcome is post-selected. The control qubit is initially state $\frac{1}{\sqrt{2}} (|0\rangle + |1\rangle)$.}
\label{2path_fom_fig}
\end{figure}

\emph{Figure of merit for measurement strategies - } At this point, is natural to wonder about the choice of the measurement outcome $|\psi_m\rangle$ on the control qubit that yields the best results. If the noise $p$ is unknown, then it makes sense to integrate over the entire noise range, and propose the following figure of merit.
\begin{equation}
\mathcal{K} (|\psi_m\rangle) =   \int_{0}^{1/3} \max \left[\mathcal{F} (|\psi_m\rangle , \rho_{\text{input}}) - \frac{2}{3},0 \right] dp 
\end{equation}
 \noindent The choice of the integrand reflects the fact that any fidelity below $2/3$ is as good as obtaining no quantum advantage at all. The optimal measurement outcome choice for the control is the one for which this quantity $\mathcal{K}$ is maximized. For the two path case, Fig. \ref{2path_fom_fig} depicts the situation. It is observed that the best measurement outcome is $|+\rangle$, that is, the same state in which the control qubit was initialized. 
 
\begin{figure}
\includegraphics[scale=0.35]{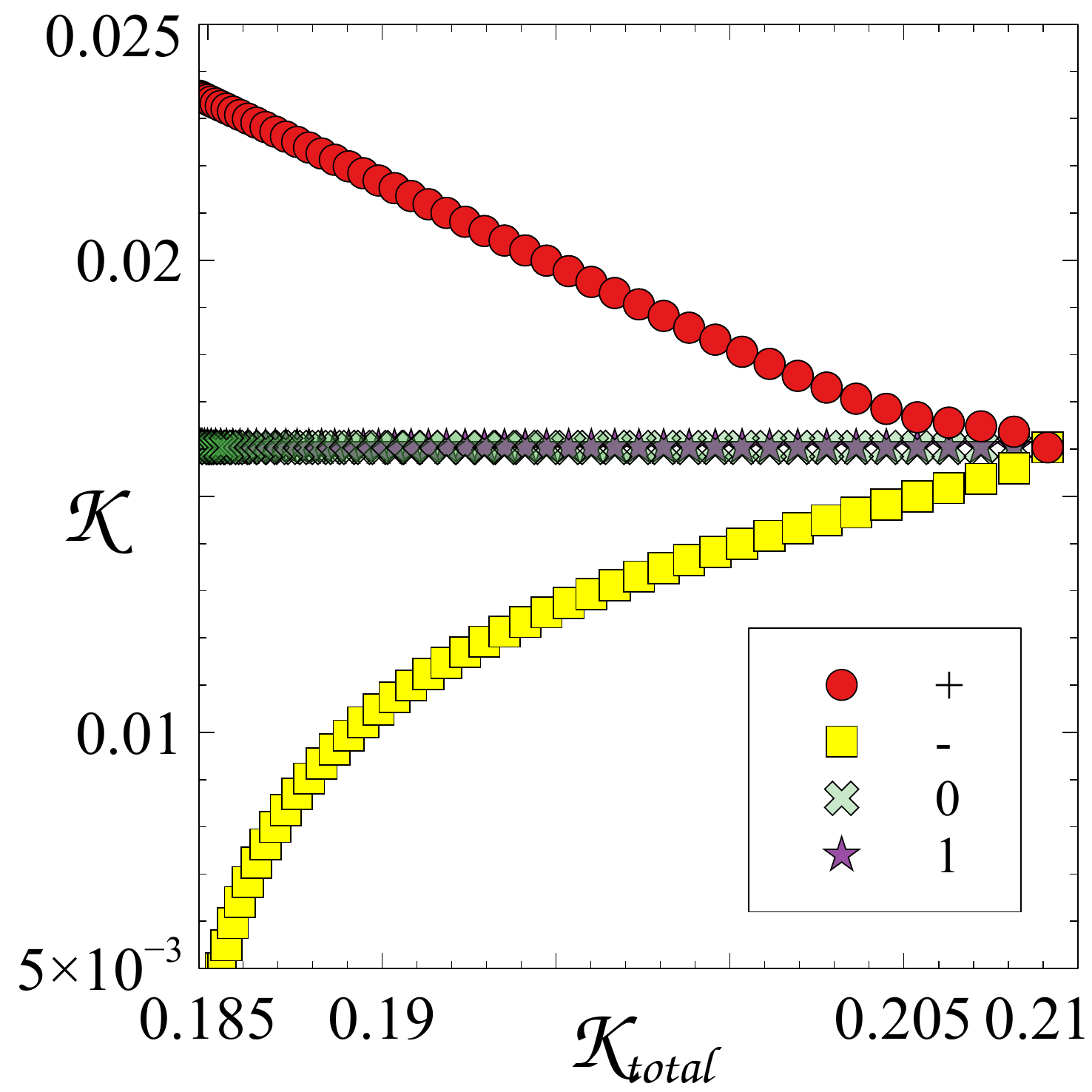}
\caption{Figure of merit $\mathcal{K}$ vs. total fidelity $\mathcal{K}_{\text{total}} = \int \mathcal{F}( \rho \otimes \rho_c, \Lambda[\rho \otimes \rho_c])$ for pure input states, and pure control states, for various outcomes (red circles for outcome $+$, yellow squares for outcome $-$, green crosses and pink stars for outcomes $0$, and $1$ respectively. As the superposition in the control state decreases, the curves converge.}
\label{tradeoff}
\end{figure}

 \emph{A tradeoff in fidelity-} Our goal in a teleportation protocol is to ensure the transmission of the qubit in question with minimal possible distortion. Interestingly, we find that the better this fidelity can be made by means of postselecting on a suitable outcome of the measurement on the control qubit, the more distant the system-control joint output state is from the input system-control uncorrelated state. This trade-off is explicitly depicted in Fig \ref{tradeoff}. 
 
\begin{figure}[h]
\includegraphics[scale=0.35]{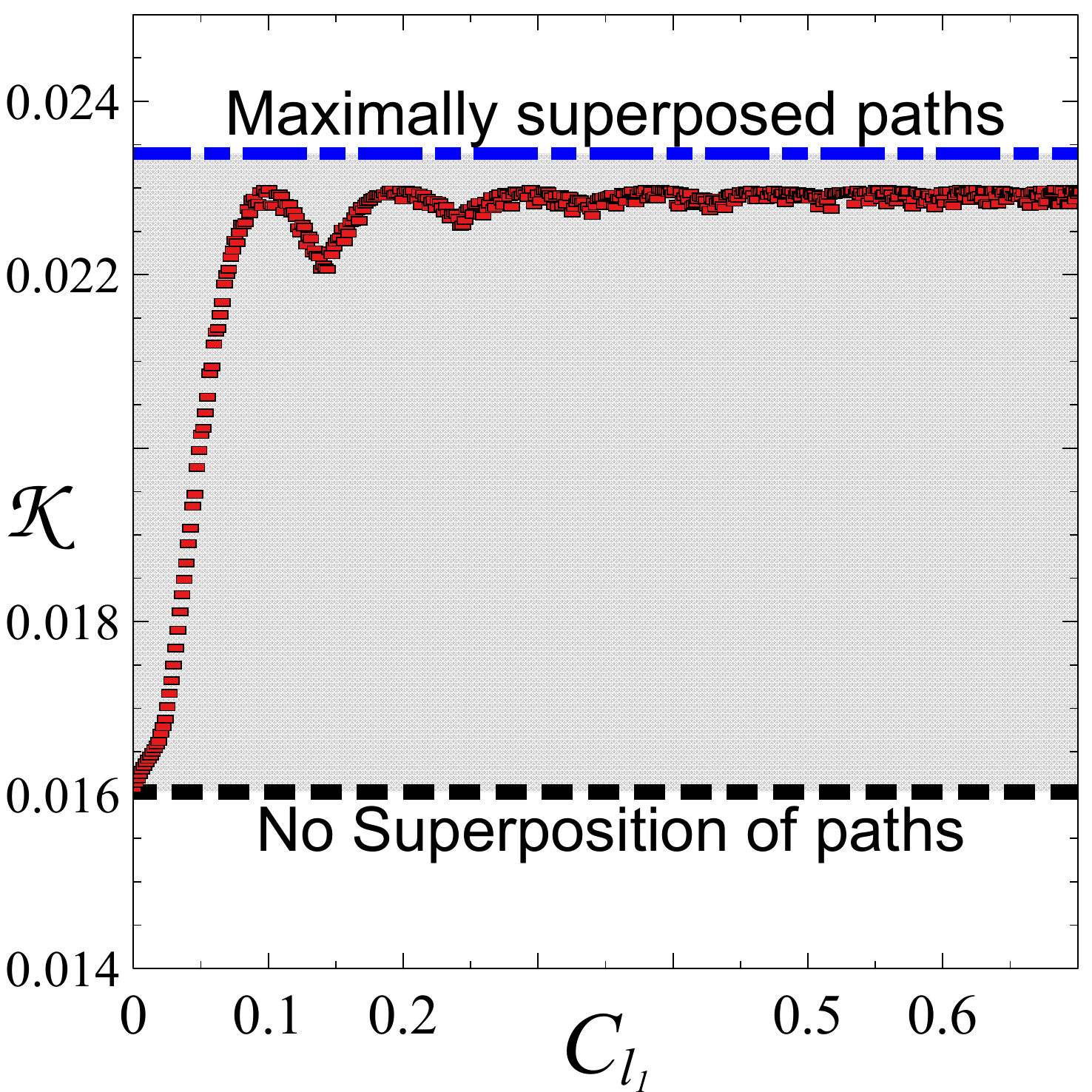}
\caption{Figure of merit $\mathcal{K}$ vs. $l_1$-norm of coherence for the initial state of the pure input control qubit, the former optimised over all possible choices of measurement outcome.}
\label{coh_fig}
\end{figure}

\emph{Link with quantum coherence -} The quantum switch superposes two distinct pathways. Hence any operational advantage gained through using a quantum switch can be  expected to be ascribed to the amount of quantum coherence \citep{baumgratz, aberg,colloquium} present in the initial state of the control qubit. However, creating a maximally coherent state may be difficult in practice. Thus, a natural question arises, namely, given the amount of coherence in the initial state of the control qubit, how much advantage is gained through the use of the quantum switch. Fig. \ref{coh_fig} indicates, that it is possible to achieve a significant boost in the figure of merit $\mathcal{K}$, even with a relatively small amount of interference between the paths. 

\emph{Multiple pathways-} Since superposing two imperfect teleportation channels provide us with enhanced teleportation fidelity, it is natural to wonder whether superposing multiple such channels would be even better. For $N$ channels, there may be $N!$ possible pathways, each associated with the now $N!$-dimensional control qubit like before. Thus, the joint output state of the qubit and the control qubit is now given by

\begin{figure}[h]
\includegraphics[scale=0.35]{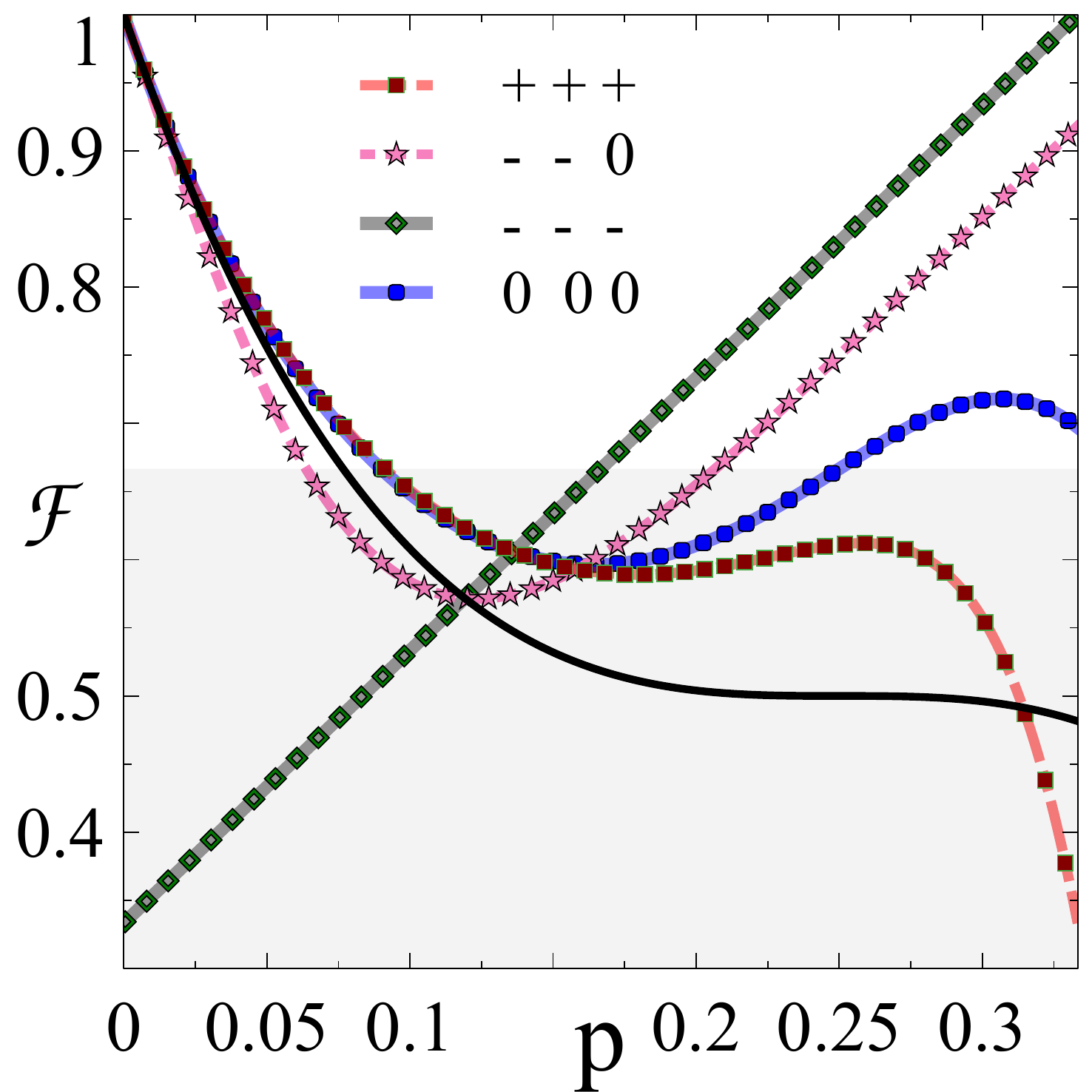}
\caption{Noise vs. teleportation fidelity profile for different measurement outcomes on the control qudit. The solid black line represents the case for three teleportation channels $A_1, A_2, A_3$ applied in succession without any quantum switch. The other lines represent a measurement outcome corresponding to (unnormalized) outcome $ |0\rangle_{A_1 A_2 A_3} + |1\rangle_{A_2 A_3 A_1} + |2\rangle_{A_3 A_1 A_2} + |\alpha_1 \rangle_{A_1 A_3 A_2} + |\alpha_2 \rangle_{A_2 A_1 A_3} + |\alpha_3 \rangle_{A_3 A_2 A_1}$ with the tuple $(\alpha_1, \alpha_2, \alpha_3) = (1,1,1)$ (red line with red rectangle points), $ (0,0,0)$ (blue line with blue rounded rectangle points), $(-1,-1,0)$ (pink line with pink star points), and  $(-1,-1,-1)$ (grey line with green diamond points).}
\label{3path_fig}
\end{figure}

\begin{widetext}
\begin{equation}
\rho_{out} =  \sum_{\pi} \sum_{i_1 i_2.... i_n}  p_{i_1} p_{i_2}...p_{i_n} \left [ \sigma_{\pi_{i_1}}\sigma_{\pi_{i_2}}...\sigma_{\pi_{i_n}} \otimes \hat{\pi} (|0\rangle \langle 0|) \right] (\rho \otimes \rho_c) \left [ \sigma_{\pi_{i_1}}\sigma_{\pi_{i_2}}...\sigma_{\pi_{i_n}} \otimes \hat{\pi} (|0\rangle \langle 0| ) \right]^{\dagger},
\end{equation}  
\end{widetext}
 where $\pi$ are the possible permutations and $\hat{\pi}$ are the corresponding basis permutation operators. The control is assumed to be  initialized in a (normalized) pure state $\sum_{i = 0}^{N! -1} \sqrt{q_{i}} |i\rangle$ . The final measurement on the control qubit is done on some basis, which is complementary to the computational basis. The analogue $\alpha$ of the two-pathway outcome $+$ , over which we post-select, now remains to be chosen. Fig. \ref{3path_fig} depicts the result for various such measurement outcome choices for three paths. It seems that post-selecting over  the control state $\sum_{i}\pm \frac{1}{\sqrt{6}}|i\rangle$, where the $+$ sign is for paths which are an even permutation of the an arbitrarily picked `original' pathway, and $-$ sign is for paths which consist of channels in odd permutations of the `original' pathway is an optimal one if the goal is to ensure high fidelity if the noise is very high. In fact this achieves perfect fidelity in the limit $p \rightarrow 1/3$. However, for low noise (unless p = 0), this choice achieves very low fidelity. 
 
In comparison, using the standard teleportation protocol thrice in succession fails to attain teleportation fidelity beyond $2/3$ unless $p < \frac{1}{4} (1- 3^{-1/3}) \approx 0.07$. From Fig. \ref{3path_fig}, one feature is quite noteworthy, the measurement choices which are good for obtaining high teleportation fidelity in the high noise regime perform relatively poorly in the low noise regime. In the case of three channels in a superposition of causal order, we consider the following family of measurement outcomes to post-select. 
\begin{equation}
\hat{M} (\phi, \lambda) = |\psi\rangle \langle \psi|, \\
|\psi\rangle = \frac{1}{N_{0}} \left( \sum |j\rangle_{even}  - \lambda \text{e}^{i \phi}  \sum |j\rangle_{odd} \right)
\label{outcome_choice_3}
\end{equation}
 
\begin{figure}
\includegraphics[scale=0.35]{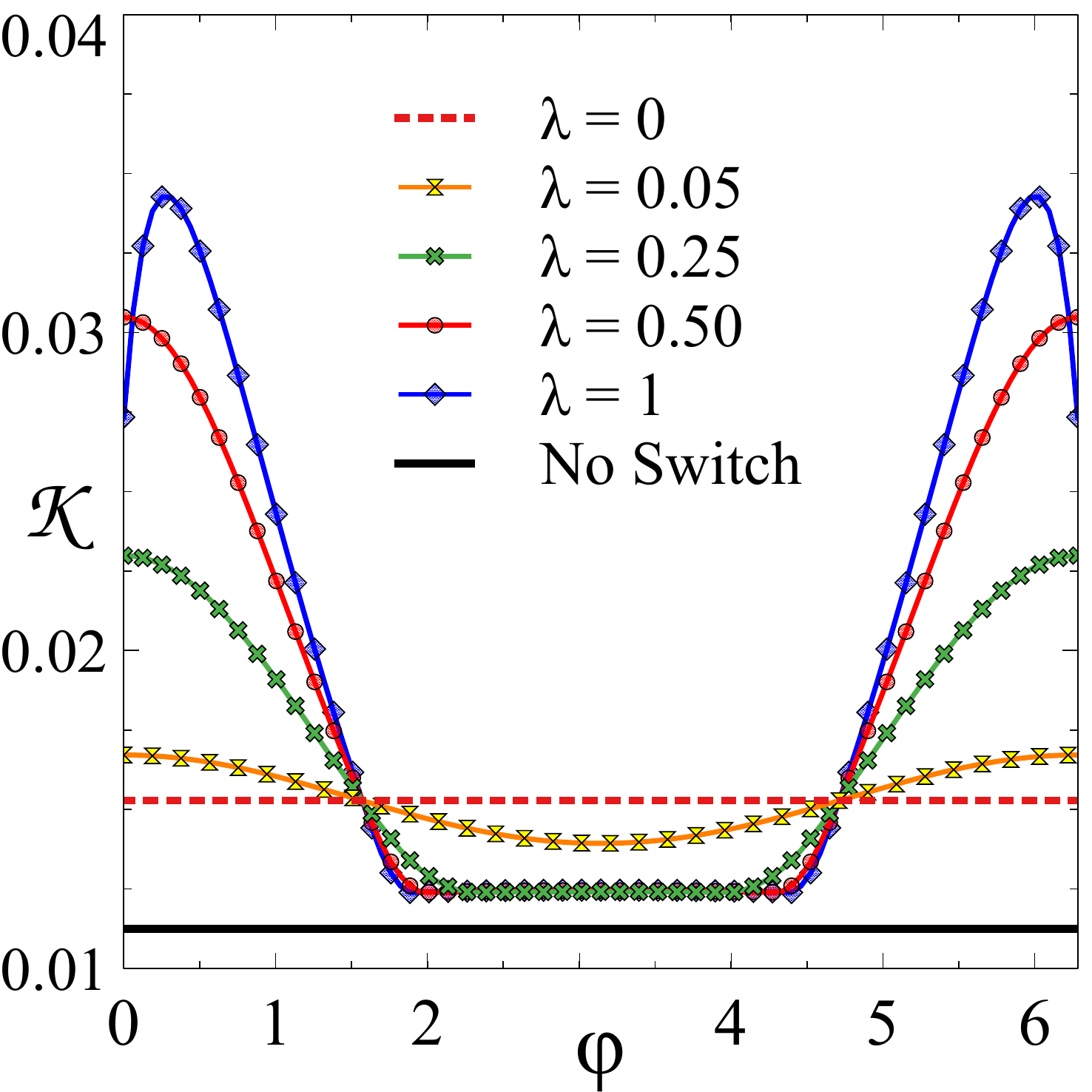}
\caption{Figure of merit $\mathcal{K}$ vs. relative phase $\phi$ between even and odd permutaton paths for varying relative weights of even and odd permutation paths of the family of measurement outcomes represented in \eqref{outcome_choice_3}. }
\label{measurement}
\end{figure}

From the Fig \ref{measurement}, we observe a few interesting features. Firstly, for all the possible measurement choices, the figure of merit lies above that of the process without any quantum switch, thus indefinite causal order qualitatively helps. Secondly, the absolute maximum is achieved at about $\phi \approx \pi/12$ for $
\lambda = 1$, unlike the two path scenario. A systematic investigation of the measurement strategy to adopt in specific cases is beyond the scope of the present work, but will be considered elsewhere. 

\emph{Conclusion-} We have shown that the presence of a quantum switch can significantly augment the teleportation fidelity under a specific noise model. The measurement on the control qubit was necessary to exploit the correlation between the control and the original qubit during the course of the evolution. In this work, we have abstained from rigorously optimizing this strategy, i.e., choosing which basis to measure and which outcome to post-select, in a general manner. This will be especially important in the multiple pathways scenario, where the available measurement choices are vastly wider than the two-pathway case. More generally, we believe that the processes with no definite causal order may turn out to be useful in other canonical quantum communication schemes as well. In the present context, it is clear that the coherence in the qubit state is leading to an operation advantage in the form of an increased teleportation fidelity. It may be interesting to investigate whether the extent of such advantage gained for specific measurement outcomes can be quantitatively proved to be a coherence monotone in the usual resource theoretic sense for more general qudit cases.  
\\

\emph{Acknowledgement-} CM acknowledges funding through a doctoral research fellowship by the Department of Atomic Energy, Government of India, as well as a fellowship by INFOSYS. We wish to thank the members of The Institute of Mathematical Sciences for their kind hospitality, where the present work was finished.

\bibliographystyle{apsrev4-1}
\bibliography{acausal_teleportation}

\end{document}